# Low temperature synthesis, magnetic and magnetotransport properties of $(La_{1-x}Lu_x)_{0.67}Ca_{0.33}MnO_3$ ($0 \leq x \leq 0.12$) system

D. Das[1], M. R. Raj[1], C. M. Srivastava[2], A. K. Nigam[3], D. Bahadur[1], and S. K. Malik[3]

[1]Department of Metallurgical Engineering and Materials Science, IIT Bombay, Mumbai 400 076, India.

[2]Department of Physics, IIT Bombay, Mumbai 400 076, India.

[3]Tata Institute of Fundamental Research, Colaba, Mumbai 400 005, India.

*Abstract*

We have been able to synthesize $Lu^{+3}$ substituted $La_{0.67}Ca_{0.33}MnO_3$ (LCMO) by an auto-combustion method. Synthesis of this compound is not successful by conventional ceramic or other chemical methods. Magnetic and electrical transport properties of the Lu substituted LCMO [$(La_{1-x}Lu_x)_{0.67}Ca_{0.33}MnO_3$ ($0 \leq x \leq 0.12$)] system have been investigated and compared with those of the $Y^{+3}$, $Pr^{+3}$, $Dy^{+3}$ and $Tb^{+3}$ substituted LCMO systems. All the compounds show a ferromagnetic metal to paramagnetic insulator transition at $T_C$. The tolerance factor reduces from 0.917 for $x = 0$ to 0.909 for $x = 0.12$ and for this range all are ferromagnetic metals indicating the dominance of the coupling between spins due to double exchange over the antiferromagnetic superexchange interaction. The transition temperatures and magnetization decrease as the Lu concentration increases. This is satisfactorily accounted for on the basis of transition from ferromagnetic at $x = 0$ to canted spin order for $x > 0$. All the samples show higher magnitude of MR compared to that in pure LCMO at 80 kOe field in the temperature range of 5 to 320K. A fairly high value of low field magnetoresistance (LFMR) of about 30% is obtained in all the samples at a field less than 5 kOe.

# 1. Introduction

Mixed valent manganites of the general formula $Ln_{1-x}A_xMnO_3$ (Ln = rare earth ions like La, Sm, Pr, etc., A = alkaline ions like Ca, Sr, Ba, etc.) have been the subject of intense research since the discovery of colossal magnetoresistance (CMR) in this class of compounds[1,2]. The parent compound, $LaMnO_3$, is an antiferromagnetic insulator in which Mn is present in a single oxidation state ($Mn^{+3}$). But substitutions by a divalent alkaline earth metal ion like Ca, Sr, Ba or Pb at the trivalent rare earth site causes mixed valency of Mn ions and transforms this compound to a ferromagnetic metal. The doped charge carriers (holes) mediate the ferromagnetic interactions between the localized spins ($t_{2g}$) associated with the variable valency Mn ions ($Mn^{+3}/Mn^{+4}$) in the crystallographically equivalent sites through the double exchange (DE) mechanism[3]. Since the ferromagnetic Curie temperature ($T_C$) and the metal insulator transition temperature ($T_{MI}$) in such compounds are related to the strength of the transfer integral between $Mn^{+3/+4}$ through the $Mn^{+3} - O - Mn^{+4}$ path, there is a strong interplay between the structural, magnetic and transport properties. The study of Hwang et al.[4] on $(La_{1-x}R_x)_{0.7}Ca_{0.3}MnO_3$ in which some $La^{+3}$ is replaced by larger size, $Pr^{+3}$ and smaller size $Y^{+3}$, reveals that the best magnetic and transport properties are obtained for a $Mn^{+3}/Mn^{+4}$ ratio of 7/3. These authors also found that the $T_C$ and the conductivity decrease with increasing $Pr^{+3}/Y^{+3}$ content, or in other words, the magnetic and transport properties of the system are a strong function of average A site ionic radius, $<r_A>$. Based on their results, they have concluded that the electronic and magnetic states are dependent on a geometrical index called the Goldsmith tolerance factor, t, which decides the crystallographic distortions from the ideal cubic (t = 1) perovskite structure. The Goldsmith tolerance factor describes the stability of the perovskite structures and is defined as $t = (r_A + r_O)/\sqrt{2}(r_B + r_O)$, where $r_A$, $r_B$ and $r_O$ are the average ionic radii of the A, B site cations and of $O^{2-}$ ion, respectively. With decreasing t, the Mn – O – Mn bond angle, to which the transfer integral is related and which describes the electron hopping between $Mn^{+3}$ and $Mn^{+4}$, decreases. In addition to the substitution by divalent cations, like Ca, Sr or Ba, the $Mn^{+3}/Mn^{+4}$ ratio is also very sensitive to the oxygen stoichiometry[5]. So the critical parameters on which the transport and magnetic properties of the manganites are strongly dependent are: (a) the average ionic radius of the A site cations, $<r_A>$, (b) the $Mn^{+3}/Mn^{+4}$ ratio and (c) the tolerance factor, t.

A considerable amount of work has been devoted to bring about the Mn – O – Mn bond angle deformation and subsequent increase of the magnetoresistance by substitutions at the rare earth site[6-9]. Application of external hydrostatic pressure[10,11] also produces similar effect in the $Mn^{+3} – O – Mn^{+4}$ network. The study by Jin et al[12] on the composition $La_{0.6}Y_{0.07}Ca_{0.33}MnO_3$ triggered a large amount of A site substitution work by light or heavy rare earth ions in the perovskite structure. Recently, it has been shown[13,14] that the increase in ionic size mismatch at the A site, with a fixed substitution level, leads to decrease in both $T_{MI}$ and magnetoresistance ratio.

A large magnetoresistance effect at 140K in $La_{0.6}Y_{0.07}Ca_{0.33}MnO_3$ system[12] made us interested to study the effect of substituting Lu at the La site of the LCMO, since the ionic sizes of Y and Lu are comparable[15] ($Y^{+3}$ = 1.075 Å and $Lu^{+3}$ = 1.032 Å) and both are diamagnetic. In spite of similar sizes of $Y^{+3}$ and $Lu^{+3}$, a difference in structural and thermal properties of the hexagonal $YMnO_3$ and $LuMnO_3$ is observed[16]. The out-of-plane lattice constant, c, continuously decreases with temperature in the range 300K to 1000K for $YMnO_3$, while it remains constant with temperature in that range for $LuMnO_3$. An entirely different variation of excess specific heat, $\Delta C_p/T$, a quantity related to the magnetic contribution of the system and obtained by the subtraction of the lattice specific heat from the net specific heat of the system, with temperature is observed[17] in the two compounds, YMnO$_3$ and LuMnO$_3$. Whereas a prominent second peak at around 50K (along with the one at around 100K) is observed in the $\Delta C_p/T$ vs. T plot in LuMnO$_3$, it is hardly found in YMnO$_3$ at the same temperature. A probable difference in orbital ordering may be a cause of these differences. Terai et al[18] have studied the electronic and magnetic properties of (La-Dy)$_{0.7}$Ca$_{0.3}$MnO$_3$ system and arrived at a phase diagram which shows the regions of spin-glass, ferromagnetic metal and paramagnetic insulating phases in a ($T_c/T_g$) vs. t diagram. They find low temperature (T ≤ 50 K) spin glass phase for t < 0.907.

The present investigation intends to study the system (La$_{1-x}$Lu$_x$)$_{0.67}$Ca$_{0.33}$MnO$_3$ (0.0 < x ≤ 0.12), with fixed Mn$^{+3}$/Mn$^{+4}$ ratio of 7/3, in terms of magnetic and electrical transport properties. Although the effect of lanthanide substitution in the A site of the perovskite LCMO or LSMO is extensively studied and documented in the literature[6-9], no report is available so far on the Lu doping in LCMO. Some workers[19,20] have reported magnetic and transport properties of Lu substituted LSMO, but those are not on single phase compositions. Instead, these compositions are solid solutions of the terminal compounds La$_{0.7}$Sr$_{0.3}$MnO$_3$ and Lu$_{0.7}$Sr$_{0.3}$MnO$_3$. The present work is the first report on the effect of this cation substitution at the A site of the perovskite LCMO. Magnetic and electrical properties of Ln substituted LCMO are compared with those of Y substitution reported in the literature[4,12]. Sushko et al.[21] and Subramanian et al.[22] have studied the effect of Lu substitution on the magnetism and colossal magnetoresistance behavior of the A$_2$Mn$_2$O$_7$ (A = Tl, In, Y, Lu) and A$_2$Mn$_2$O$_7$ (A = Dy –Lu, Y, Sc, In or Tl) pyrochlores, respectively. Sushko et al. find that applied pressure in the Lu doped pyrochlore suppresses ferromagnetism. They have suggested that in the CMR pyrochlore the nearest-neighbor superexchange couplings are antiferromagnetic in nature, while ferromagnetism originates from the long range interactions dominated by Mn 3d – Mn 3d superexchange via Tl 6s orbitals. Subramanian et al. have observed an increase in resistivity of the pyrochlore with Lu substitution. They have suggested a spin-glass like behavior from the susceptibility and powder neutron diffraction studies. Terai et al[18] have studied the electronic and magnetic properties of $(La-Dy)_{0.7}Ca_{0.3}MnO_3$ and compared them with compounds containing $Y^{3+}$, $Pr^{3+}$ and $Tb^{3+}$. They find that for t < 0.907, a spin-glass insulating state appears at low temperatures. This insulating state is different from the ferromagnetic insulator in the compounds containing $Y^{3+}$ and $Pr^{3+}$. Since in our case, the ionic radii of $Y^{3+}$ and $Lu^{3+}$ are nearly the same, the tolerance factor t exceeds 0.907 for all values of $0 \leq x \leq 0.12$ and, like in $(La-Y)_{0.7}Ca_{0.3}MnO_3$, the present $(La_{1-x}Lu_x)_{0.7}Ca_{0.3}MnO_3$ system is a ferromagnetic metal for all values of x. This is accounted for on the basis of a two sublattice model in which the spins interact through the double exchange and superexchange interactions. It is shown that, depending on the competition between the ferromagnetic double exchange and the antiferromagnetic superexchange, the ground state at low temperatures could be ferromagnetic-canted spin metal (FMM), antiferromagnetic insulator (AFI) or spin glass insulator (SGI).

## 2. Experimental

The $(La_{1-x}Lu_x)_{0.67}Ca_{0.33}MnO_3$ samples with x = 0.0 to 0.12 were synthesized using the auto-combustion method[23]. Stoichiometric amount of $La_2O_3$, $Lu_2O_3$, $CaCO_3$ and Mn-acetate tetrahydrate (Aldrich chemicals with purity better than 99.9%) were dissolved in minimum amount of distilled water. Concentrated $HNO_3$ was added to this cationic stock solution to convert all the metal ions into respective nitrates followed by heating at around $60^0C$ for 10 minutes on a hot plate. Glycine was added to this transparent aqueous nitrate solution maintaining a glycine/nitrates molar ratio of 0.5. Subsequent heat treatment of this nitrate-glycine mixture at around $80^0C$ on the hot plate led to the thermal dehydration of the excess solvent thereby producing a thick viscous liquid, the manganite precursor. As soon as the viscous liquid formed, the temperature of the hot plate was raised to $200^0$C. The viscous liquid swelled up and auto ignited with rapid evolution of huge quantity of gases to produce voluminous powders. The heat liberated in the strong exothermic reaction between the fuel (glycine) and the oxidant (the nitrate mixture) assists in forming the phase. The nature of the powder so formed is sensitive to this fuel to oxidant ratio[23]. The room temperature powder X-ray diffraction pattern of the as obtained autoignited powder showed the formation of the compound with somewhat broad reflection lines, which suggested improper crystallization of the perovskite phase. To drive off the excess undecomposed glycine, nitrates and their decomposition products from the system and to obtain the desired phase in proper form, the auto ignited powders were subjected to calcinations at $600^0$ C for 2 hrs in a muffle furnace. The cold pressing of the calcined powders in a uniaxial hydraulic press with subsequent sintering of the green pellets at $800^0$ C for 2 hrs in static air to the formation of the bulk polycrystalline $(La_{1-x}Lu_x)_{0.67}Ca_{0.33}MnO_3$ samples.

Desired phase formation in the $(La_{1-x}Lu_x)_{0.67}Ca_{0.33}MnO_3$ samples was confirmed by the room temperature X-ray powder diffractions of the final sintered powders in a Philips X'Pert Diffractometer (model PW 3040/60) using $CuK_\alpha$ radiation and a continuous scan from $20^0 - 80^0$ with a scanning rate of $0.02^0$/15 secs. Magnetization (M) measurements as a function of temperature (T) and applied magnetic field (H) (M vs. T and M vs. H) were carried out using a SQUID magnetometer (MPMS 7, Quantum Design) in the temperature range of 5 to 300K. The temperature dependence of electrical resistivity, $\rho(T)$, and the field response of the electrical resistivity, $\rho(H)$, were measured by a standard four-probe dc technique using a commercial set-up (PPMS Model 6000,

Quantum Design) in the temperature range of 5 to 320 K. The magnetoresistance of the samples was measured at 80 kOe field in the same temperature range.

## 3. Results and Discussions

The formation of single phase Lu substituted LCMO compounds is a sensitive function of heat treatment temperature and the synthesis process. The conventional solid state and the citrate gel routes have not been successful in yielding single phase compounds with Lu substitution. Moreover, raising the heat treatment temperature favors the formation of more stable $LuMnO_3$. To overcome these difficulties, we have adopted an auto-combustion process with the final sintering protocol being kept at $800^0C$ for 2 hrs. This was necessary as the precursor decomposes to phases such as $LuMnO_3$ at higher temperatures. It is noteworthy that this kind of decomposition occurs only in the case of Lu substitution. Room temperature powder x-ray diffraction patterns of the samples are shown in Fig. 1. The patterns show an orthorhombic symmetry with shifting of the reflection lines indicating that Lu has gone into the LCMO lattice. Absence of any extra lines confirms the phase purity of the samples. The shifting of the most intense line (002, 200) on Lu substitution is shown in the inset of Fig. 1. The continuous decrease of the unit cell lattice volume, shown in Table 1 (determined from the least-squares fitting of the diffraction pattern) confirms the incorporation of the smaller size Lu at the La site of LCMO. The particle size, estimated from X-Ray line broadening using Scherrer's formula is found to be around 50 nm for all the compositions. This is also supported by our direct observations using transmission electron microscopy (TEM). TEM pictures for $x = 0$ and $x = 0.10$ are given in Fig. 2 (a) and (b), respectively. The micrographs show that the particle sizes are in the range of 48 – 55 nm for different samples.

The magnetization behavior of the $(La_{1-x}Lu_x)_{0.67}Ca_{0.33}MnO_3$ samples, as in the form of reduced magnetization, m (= M(T)/M(0)) vs. reduced temperature, t (= $T/T_C$) is shown in Fig. 3 in the temperature range of 5 to 300K in 5 kOe field. The magnetization behavior of all the samples agrees reasonably well with a simple Heisenberg model for S = ½ as shown as a solid line in the same temperature range. All the samples show a paramagnetic to ferromagnetic transition, $T_C$, defined from the maximum inflexion of the Zero Field Cooled (ZFC) magnetization data and this decreases from 258K for x = 0 to 220K for x = 0.12. Also, the saturation magnetization decreases as x increases. The isothermal saturation magnetization, $M_S$, at 5K measured in 50 kOe field inset (a) of Fig. 3) decreases from 76.9 emu/gm for x = 0 to 58.1 emu/gm for x = 0.12 (inset (b) of Fig. 3). The values of $T_C$ and $M_S$ for all the samples are given in Table 2. The interesting point to note is that the transition width, $\Delta T$, (defined by the difference between 90% and 10% of the saturation magnetization at 0 K, $M_s(0)$), in the magnetization curve (M vs. T), increases as x increases. For x = 0, $\Delta T$ is 150K which increases to 171K for x = 0.12. For S = 1/2, $\Delta T \sim 0.4 T_C$ (= 104K). Similar observations have been made in Y substituted samples[24,25]. Freitas et al.[24] have also observed a decrease in $T_C$ in the $(La_{1-x}Y_x)_{0.7}Ca_{0.3}MnO_3$ samples from 252K for x = 0 to 89K for x = 0.15, calculated from the minimum in the dM/dT vs. T behavior of the field-cooled (FC) magnetization measured at 50 Oe. When Lu substitutes for La in the LCMO lattice, the ferromagnetic interactions weaken and this weakening increases with increasing Lu content because of the increased distortion of the $Mn^{+3} - O - Mn^{+4}$ bond angle and decreased Mn – O bond distances[26,27]. It is necessary to understand the correlation between the magnetic structure and the crystal structure as reported by Terai et al[18] in the form of phase diagram relating the tolerance factor, t, to $T_C$. The magnetic ordering at low temperature shows that in (La-R)$_{0.7}$Ca$_{0.3}$MnO$_3$ (R = Dy, Tb, Pr, Y) a triple point is seen at the tolerance factor, t = 0.908 and for T < 50K. We expect the present system to follow this phase diagram as $Lu^{+3}$ has similar ionic radius since $Y^{+3}$ and the tolerance factors in the present case lie between 0.909 to 0.917. Hence all presently studied compositions should be Ferromagnetic Metal (FMM). This phase diagram can be understood on the basis of competition between the superexchange and the double exchange interactions between spins on a general two sublattice model. We take that the superexchange interaction is isotropic and is given by $E_s = - J\, S_1.S_2$ and the double exchange is given by $E_d = - \varepsilon_p \xi \cos(\theta/2)$, where $\theta$ denotes the angle between the spins on sublattices 1 and 2, $\varepsilon_p$ denotes the gain in energy for transfer of parallel spins in nearest neighbor $e_g$ orbitals for the double exchange interactions and $\xi$ is the fraction of $Mn^{+4}$ ions on the B-site. If there are z nearest neighbors on sublattice 2 for a spin on sublattice 1, the interaction energy can be written as[28],

$$U = -\varepsilon_p \xi\, Nz \cos(\theta/2) + |J| S^2 Nz \cos\theta \quad\ldots\ldots\ldots\ldots\ldots\ldots\ldots\ldots\ldots\ldots(1)$$

where, N is the number of magnetic ions per unit volume. Minimizing U with respect to $\theta$ gives two solutions:

(i)      $\sin(\theta/2) = 0$           for $\varepsilon_p \xi > 4|J| S^2$    ............................(2)

and (ii)      $\cos(\theta/2) = \varepsilon_p \xi / 4|J| S^2$      for $\varepsilon_p \xi < 4|J| S^2$    ....................(3)

For (i), $\theta = 0$ and FMM phase is obtained and for (ii) $\theta$ lies between 0 and $\pi$ and a canted spin order or an antiferromagnetic spin order is obtained. When the interaction energy due to double exchange in eq. (1) is nearly equal to the superexchange energy, the spin glass behavior is expected. The value of J depends on the bond angle Mn – O – Mn and hence on the tolerance factor and, as t decreases, the bond angle deviates more from $180^0$ and $|J|$ decreases. When $\varepsilon_p\xi$ becomes comparable to $4JS^2$, the FMM phase changes to SGI phase as shown by Terai et al[18]. Similar type of conclusion from the mean field theory of magnets with competing double exchange and superexchange interactions has been reached by Golosov et al[29].

In the present system for $x = 0$, $\xi = 0.33$, the spin order is a well known collinear ferromagnet. In general, spins are distributed amongst four sublattices[30]. However, if each Mn ion has an average charge and spin lying between +3 and +4, and 3/2 and 2, respectively, the superexchange interaction can be described by a single exchange constant J as in eq. (1) and a two sublattice model can be used. The eight magnetic ions in a magnetic unit cell of cell constant 2a occupying the sites (000) and (aa0) constitute one sublattice and those occupying (a00) and (aaa) constitute the second sublattice, and solutions (i) and (ii) obtained on the basis of the two sublattice hold. The sublattice magnetization will follow the Brillouin expression. For the simple case of spin ½, the reduced magnetization $m = M(T)/M(0)$ is related to the reduced temperature $t = T/T_C$ as

$$m = \tanh(m/t) \quad\quad\quad\quad\quad\quad\quad\quad\quad\quad\quad\quad (4)$$

If the magnetization of each of the sublattices is equal, the net magnetization is

$$\overline{m} = m\cos(\theta/2) \quad\quad\quad\quad\quad\quad\quad\quad\quad\quad\quad\quad (5)$$

In Fig. 3, we have plotted the curve calculated using eq. (4) and compared it with the observed temperature dependence of magnetization for $La_{0.67-x}Lu_xCa_{0.33}MnO_3$ ($x = 0$, 0.07, 0.10, 0.12) samples. For $x = 0$ sample, the magnetization curve is closest to the curve given by eq. (4). As x increases, the curve departs increasingly from the curve with $x = 0$. This is also shown in the inset (a) of Fig. 3 in which M is plotted as a function of H at 5K. The magnetization at 5K and in 50 kOe field decreases linearly with x as shown in the inset (b) due to the presence of canting. This follows because with the introduction of Lu, the number of charge carriers decreases as $\xi(1-x)$; so from eq. (3)

$$\overline{m} = m\left[\frac{\varepsilon_p \xi(1-x)}{4JS^2}\right] \quad\quad\quad\quad\quad\quad\quad\quad\quad\quad (6)$$

This accounts for the linear decrease of effective magnetic moment $\mu_{eff}$ per Mn atom, with x as given in Table 2 and shown in Fig. 4. The observed gradient, $d\overline{m}/mdx$ is -1.85. This may be compared with the theoretical value -1.0 obtained from eq. (6) when reasonable values of $\varepsilon_p$ = 260K, $\xi$ = 0.33, J = 9.5K and S = 1.5, as discussed in ref. 30, are taken. The deviation from theory is due to the low field (50 kOe) used here which does not lead to saturation. Similar M vs. H behavior is observed for $La_{2/3-x}Y_xCa_{1/3}MnO_3$ (x = 0 to 0.25) by Fontcuberta et al[32]. With these parameters the canting angle $\theta$ is obtained from eq. (3) and is plotted in Fig. 4 along with the values of $\theta$ obtained from experiment.

The mean field theory for one sublattice relates the Curie temperature to $zJS^2$ as $k_BT_C/zJS^2$ = 2/3. A similar result has been obtained by de Gennes for the LCMO considering both the double and superexchange energies this can be expressed as[28]

$$k_BT_C = (-\frac{2}{3}zJS^2) + (\frac{2}{5})(\varepsilon_p \xi z)$$

The two constant prefactors differ due to the anisotropy of the interactions assumed in this model. In the two sublattice model, with the isotropic interactions comprising of the double and superexchange energies, an expression for $T_C$ can be obtained by equating 2/3U in eq. (1) to $k_BT_C$ for the canted to ferromagnetic structures. We then obtain

$$k_B T_C = \frac{2}{3} z \left[ \varepsilon_p \xi (1-x) \cos(\theta/2) - JS^2 \cos\theta \right] \ldots\ldots\ldots\ldots\ldots\ldots\ldots\ldots\ldots\ldots(7)$$

where $\theta$ is given by eq. (3) - With z = 6, and using the parameters $\varepsilon_p$, $\xi$, J and S given earlier to obtain $\theta$, $T_C$ has been calculated using eq. (7) and is given in Table 2 and shown in Fig. 5 along with the experimental values. The good fit to theory shows that the ferromagnetic and canted model describes satisfactorily the magnetic behavior of the Lu substituted LCMO in the range $0 \leq x \leq 0.12$.

If the A site cations are widely different in their ionic sizes and the bond angle is severely deformed so that the superexchange energy becomes comparable or dominates the double exchange energy, then the spin glass type of behavior is obtained as found for t < 0.907 in the phase diagram of Terai et al[18]. To demonstrate that there is no spin glass type of order in the present case, we have measured the in-phase ($\chi'$) and out-of-phase ($\chi''$) low field (at 11 Oe) ac susceptibility of the sample with x = 0.07 as a function of temperature at frequencies of 9.50 Hz, 95 Hz and 950 Hz. The results are plotted in Fig. 6 (a) and (b) in the temperature range of 5 to 300K. The low field magnetization of the same sample with x = 0.07 was measured during warming under 110 Oe field, after cooling down from room temperature without applied field (zero field cooled, ZFC) and subsequently under applied field (field-cooled, FC) and the results are shown in the inset of Fig. 6 (a). The in-phase susceptibility ($\chi'$) of the x = 0.07 sample, shows a maximum at a temperature which does not shift with increasing frequency. This shows that the signature for spin glass behavior is absent in the present system.

The temperature dependence of the electrical resistivity of the $(La_{1-x}Lu_x)_{0.67}Ca_{0.33}MnO_3$ samples is shown in Fig. 7 where normalized resistance [R (T)/R (T = 300K)] is plotted against temperature in the temperature range of 5 to 320K.

Following features are worth emphasizing (a) the trend in the electrical resistance variation with temperature is the same for all the samples. However, the resistance values increase with increasing x reflecting decrease in the number of charge carriers in the FMM-canted spin region. (b) all the samples show a distinct metal-insulator transition at $T_{MI}$ close to the magnetic transition temperature, $T_C$, and $T_{MI}$ shifts monotonically with x towards lower temperatures with decreasing average A site ionic radius, $<r_A>$. (c) the presence of a distinct low temperature upturn is observed in all the samples showing localization of the carriers at these temperatures. The decreasing $T_{MI}$, $T_C$ and increasing resistivity with decreasing $<r_A>$ have been also noticed in Y substituted LCMO by many authors[4,13,24,34]. When smaller size $Lu^{+3}$ (compared to $La^{+3}$ ion) substitutes La at the A site of the perovskite, the $Mn^{+3} - O - Mn^{+4}$ bond angle gets further reduced from $180^0$. Hence, the strength of the double exchange and the ferromagnetic interaction between the neighboring $Mn^{+3/+4}$ ions weakens as (1-x) in eq. (7). This is reflected in the lowering of $T_C$ and $T_{MI}$ with subsequent increase of the resistivity. The increase in normalized peak resistivity $[R_{MI}/R (T = 300K)]$ with composition, x, is shown in the inset of Fig. 7. The $T_{MI}$ values for all the compositions are given in Table 2; $T_{MI}$ shifts from 226K for x = 0.0 to 153K for x = 0.12. The reduced $T_{MI}$ and the increased peak resistivity (34.4 Ω-cm) of the pure LCMO are due to the smaller particle size (d ~ 45 nm) of the sample. Similar results have been obtained by others.[35] Although $T_{MI}$ and $T_C$ for the Y substituted samples are lower than those of the present Lu substituted samples, the variation with x essentially remains similar. The Lu substitution at the La site reduces the average A site ionic radius from 1.204 Å for x = 0.0 to 1.182 Å for x = 0.12. The gradual lattice distortion with decreasing Mn – O – Mn bond angle is also reflected in the corresponding tolerance factor which reduces from 0.917 (x = 0.0) to 0.909 (x = 0.12). The variation of $T_{MI}$ and $T_C$ with average A site ionic radius, $<r_A>$ is plotted in Fig. 8 and the expected trend is obtained which agrees well with that previously reported[27,34]. The monotonic decrease of $<r_A>$ and t with Lu concentration, x, is shown in the inset (a) of Fig. 8. In inset (b) we have plotted $T_C$-t phase diagram. As shown by Terai et al[18], for t > 0.907 the magnetic and electronic phase for $(La_{0.7-x}M_x)Ca_{0.3}MnO_3$ (M = Tb, Y, Dy and Pr) is FMM. In the present case also all the samples are metallic and ferromagnetic/canted spin since t > 0.907. The only difference is that in the present case $T_C$ drops from 258K to 220K as t changes from 0.917 to 0.909 while for Tb, Y, Dy and Pr, it drops from 258K to 80K for the same change in t. This is shown in the inset (b) of Fig. 8 in which the present data is plotted on the $T_C$-t phase diagram of Terai et al[18].

Fig. 9 shows magnetoresistance behavior of the $(La_{1-x}Lu_x)_{0.67}Ca_{0.33}MnO_3$ samples in the temperature range of 5 to 320K in an applied field of 80 kOe. All the samples show similar kind of behavior throughout the temperature range studied with a distinct peak close to the magnetic transition temperature and a low temperature rising part typical of grain boundary effect in polycrystalline samples[36]. Also the magnitude of MR increases from 53.8% to 76.3% as x increase from 0 to 0.12. The gradual increase of the peak MR with Lu concentration is shown in the inset (a) of Fig. 9 and the values of peak MR for the other samples are given in Table 2. The gradual increase in MR with progressive Lu substitution at La site is expected due to continuous lowering of $<r_A>$ and the increased distortion of the Mn – O – Mn bond angle. Similar effects in transition element substituted (in B site) LCMO have been observed earlier[37].

To study the isothermal field response of magnetoresistance, we measured MR versus field at 5K in the field range of 0 to 90 kOe and the results are shown in the inset (b) of Fig. 9. The following features of the curves are noteworthy. There are two distinct slopes in all the curves, one low field (< 5 kOe) steep slope which is nearly same for all the composition and the other is an extended high field slope which increases with increasing x. Almost 30% rise in MR occurs below 5 kOe field and then MR increases slowly with field. This field response is somewhat different from what has been observed in Y substituted LCMO samples by Damay et al.[27] and Fontcuberta et al.[34] In those cases, the low field slope also varies with composition and both the low field and the high field slopes increase with increasing Y content. However, the resistivity decreases at all measured temperatures with increasing field. This has been ascribed to the strong spin disorder inside the grains by Damay et al.[27] But Fontcuberta et al.[34] have described this low field slope (Low field MR, LFMR) in terms of interface magnetoresistance which becomes prominent for samples having narrower bandwidth. The monotonic enhancement of the LFMR is explained on the basis of the progressive reinforcement of the AF/F competition with increasing Y content which broadens the effective interface. Such type of field response has also seen in polycrystalline LSMO by many authors[38,39]. This effect has been accounted for on the basis of magnetic surface effects due to smaller particle size. Similar effects have been attributed to arise from the motion of the domain walls at low fields and the domain wall rotation at high fields[40]. In the domain boundary region, the electron hoping between $Mn^{+3}$ and $Mn^{+4}$ does not occur as readily as it does within the domain itself which accounts for the high resistivity in the low field region where the domain wall density is high. As the field increases, the number of domain walls reduces resulting in a drop in resistivity. The field where the slope changes in the normalized resistivity vs. H curve corresponds to the near-saturation field in the magnetization curve where the sample consists primarily of a single domain and subsequent changes in magnetization are due to the domain rotation. This results in much smaller changes in the magnetoresistance. In the present work also since the particle size is very small (~ 50 nm), the magnetic surface effects come into picture and the low field response is probably due to enhanced grain boundary scattering of the polarized electrons. To demonstrate this idea we have plotted the variation of magnetization and MR with the applied field in the same graph in Fig. 10 for a single composition, x = 0.12. The magnetization does not saturate in 50 kOe field. At this field the magnetic moment per Mn ion is 2.22 $\mu_B$. For this value of field for x = 0 we obtain the calculated value as 2.88 $\mu_B$ (Table 2). The steep rise of M near H = 0 indicates the presence of reversible domain wall displacement as discussed by Terai et al[18] for the $(La-Dy)_{0.7}Ca_{0.3}MnO_3$.

## 4. Conclusions

We have been able to synthesize polycrystalline single phase $(La_{1-x}Lu_x)_{0.67}Ca_{0.33}MnO_3$ (x = 0 to 0.12) samples by auto-combustion method. Synthesis of these compounds has not been possible so far by other methods e.g. ceramic or citrate gel. The particle size of all the sample is small in the range of 50 nm. Magnetic and electrical transport behavior of these samples has been studied. The substitution of smaller size Lu at the La site shifts both the metal-insulator transition temperature and the Curie temperature to lower values with a subsequent increase in electrical resistivity. A two sublattice model for magnetic behavior accounts for the observed decrease of the critical temperature and magnetization as the La concentration is increased. The decrease in average A site ionic radius from 1.204 Å for x = 0 to 1.182 Å for x = 0.12 results in change of tolerance factor from 0.917 to 0.909 which leads to an enhancement of MR from 54% to 76% in 80 kOe field.

## Acknowledgement

Financial support from Department of Science and Technology, (DST) Govt. of India, to IIT, Bombay for this work is gratefully acknowledged.

## Table Captions

Table1: Lattice parameters and unit cell volume of $(La_{1-x}Lu_x)_{0.67}Ca_{0.33}MnO_3$ obtained from the least-squares fitting of the observed $2\theta$ or d values. Particle size of the $(La_{1-x}Lu_x)_{0.67}Ca_{0.33}MnO_3$ samples has been calculated using Scherrer's formula. Tolerance factor of the samples has been calculated using the ionic radii from ref. 15.

Table2: Magnetic and electrical transport parameters of $(La_{1-x}Lu_x)_{0.67}Ca_{0.33}MnO_3$, ($0 \leq x \leq 0.12$) samples obtained from experiments. $T_C$ is the critical temperature, $M_S$ is the magnetization measured in 50 kOe at 5K, $T_{MI}$ and $\rho_{MI}$ are measured from the peak in the resistivity ($\rho$) vs. temperature (T). The calculated canting angle and $T_C$ are obtained from eqs. (3) and (6), respectively, using $\varepsilon_p = 260K$, $J = 9.5K$, $\xi = 0.33$ and $S = 3/2$ (see text).

## Figure Captions:

Fig. 1: Room temperature powder X-ray diffraction patterns of the $(La_{1-x}Lu_x)_{0.67}Ca_{0.33}MnO_3$ samples. Inset shows the shifting of the most intense line (002, 200) as x varies from 0 to 0.12.

Fig. 2: Transmission electron micrograph of $(La_{1-x}Lu_x)_{0.67}Ca_{0.33}MnO_3$ samples for (a) x = 0 and (b) x = 0.10. The micrograph shows the average particle size to be ~ 50 nm.

Fig. 3: Behavior of reduced magnetization (M(T)/M(O)) as a function of reduced temperature (T/Tc) of $(La_{1-x}Lu_x)_{0.67}Ca_{0.33}MnO_3$ ($0 \leq x \leq 0.12$) samples in 5 kOe field and in the temperature range of 5 – 300K. The solid line in the main panel is the fit for Heisenberg model with S = 1/2. Inset (a): Magnetization behavior at 5K in field upto 50 kOe. Inset (b): Variation of saturation magnetization, $M_S$, with Lu concentration in the samples at 5K and 50 kOe field.

Fig. 4: Variation of effective magnetic moment per Mn atom, $\mu_{eff}$, and canting angle, $\theta$, between spins with Lu concentration, x, in $(La_{1-x}Lu_x)_{0.67}Ca_{0.33}MnO_3$. The solid line is the $\theta$ calculated using eqs. (3) with $\varepsilon_p$ = 260K, $\xi$ = 0.33, J = 9.5K and S = 1.5 (see text).

Fig. 5: Comparison of the theoretical $T_C$ obtained using eq. (6) with experiment for $(La_{1-x}Lu_x)_{0.67}Ca_{0.33}MnO_3$ samples with $0 \leq x \leq 0.12$. The values of the parameters used are $\varepsilon_p$ = 260K, J = 9.5K, $\xi$ = 0.33 (1-x) and S = 3/2.

Fig. 6 (a): In-phase ac susceptibility ($\chi'$) behavior of the $(La_{1-x}Lu_x)_{0.67}Ca_{0.33}MnO_3$ sample with x = 0.07, measured with in ac field of 11 Oe and at frequencies 9.50, 95 and 950 Hz. Inset: ZFC and FC magnetization behavior as a function of temperature (T) for the same sample in a field of 110 Oe.

Fig. 6 (b): Out-of-phase ac susceptibility ($\chi''$) behavior of the $(La_{1-x}Lu_x)_{0.67}Ca_{0.33}MnO_3$ sample with x = 0.07, measured in an ac field of 11 Oe and at frequencies 9.50, 95 and 950 Hz.

Fig. 7: Temperature dependence of the normalized resistance, (R(T)/R(300K)), of the $(La_{1-x}Lu_x)_{0.67}Ca_{0.33}MnO_3$ ($0 \leq x \leq 0.12$) samples in the temperature range 5 – 325K. Inset: variation of the peak normalized resistivity with composition, x.

Fig. 8: Variation of metal insulator transition temperature, $T_{MI}$ and $T_C$ with $<r_A>$ for $(La_{1-x}Lu_x)_{0.67}Ca_{0.33}MnO_3$ (O) samples. The solid lines in the main panel are guide to the eye. Inset (a): Variation of average A site ionic radius, $<r_A>$, and the tolerance factor (t) with composition, x. (b): Variation of $T_C$ with t for the present case along with that obtained in ref. (18) by Terai et al for $Y^{+3}$, $Dy^{+3}$, $Pr^{+3}$ and $Tb^{+3}$ substituted LCMO systems.

Fig. 9: Variation of MR with temperature in 80 kOe for $(La_{1-x}Lu_x)_{0.67}Ca_{0.33}MnO_3$ ($0 \leq x \leq 0.12$) samples. Inset (a): Variation of peak MR% with Lu concentration, x. Inset (b): Field response of MR of the samples at 5K in applied field upto 90 kOe.

Fig. 10: Comparison of the variation of magnetization, M, and %MR with field upto 50 kOe at 5K for $(La_{1-x}Lu_x)_{0.67}Ca_{0.33}MnO_3$ sample with x = 0.12.

**Table 1**

| Lu Conc. x | a (Å) | b (Å) | c (Å) | Unit Cell Vol. (Å$^3$) | Particle size from XRD(nm) | t |
|---|---|---|---|---|---|---|
| 0.0 | 5.44 | 7.68 | 5.46 | 227.88 | 45 | 0.917 |
| 0.07 | 5.43 | 7.68 | 5.46 | 227.80 | 46 | 0.913 |
| 0.10 | 5.44 | 7.66 | 5.45 | 227.35 | 46 | 0.911 |
| 0.12 | 5.41 | 7.67 | 5.46 | 226.57 | 47 | 0.909 |

**Table 2**

| X | T$_C$ (K) | | M$_S$ (emu/g) | µ$_{eff.}$ (µ$_B$) | θ (deg.) | | T$_{MI}$ (K) | ρ$_{MI}$ (Ω cm) | Peak MR (%) |
|---|---|---|---|---|---|---|---|---|---|
|  | exp | cal |  |  | exp | cal |  |  |  |
| 0.0 | 258 | 259 | 76.9 | 2.88 | 0 | 0 | 226 | 34.4 | 53.8 |
| 0.07 | 234 | 233 | 66.0 | 2.50 | 60 | 42 | 189 | 100.3 | 67.0 |
| 0.10 | 222 | 224 | 60.8 | 2.32 | 73 | 51 | 165 | 190.0 | 73.4 |
| 0.12 | 220 | 219 | 58.1 | 2.22 | 79 | 56 | 153 | 359.5 | 76.3 |

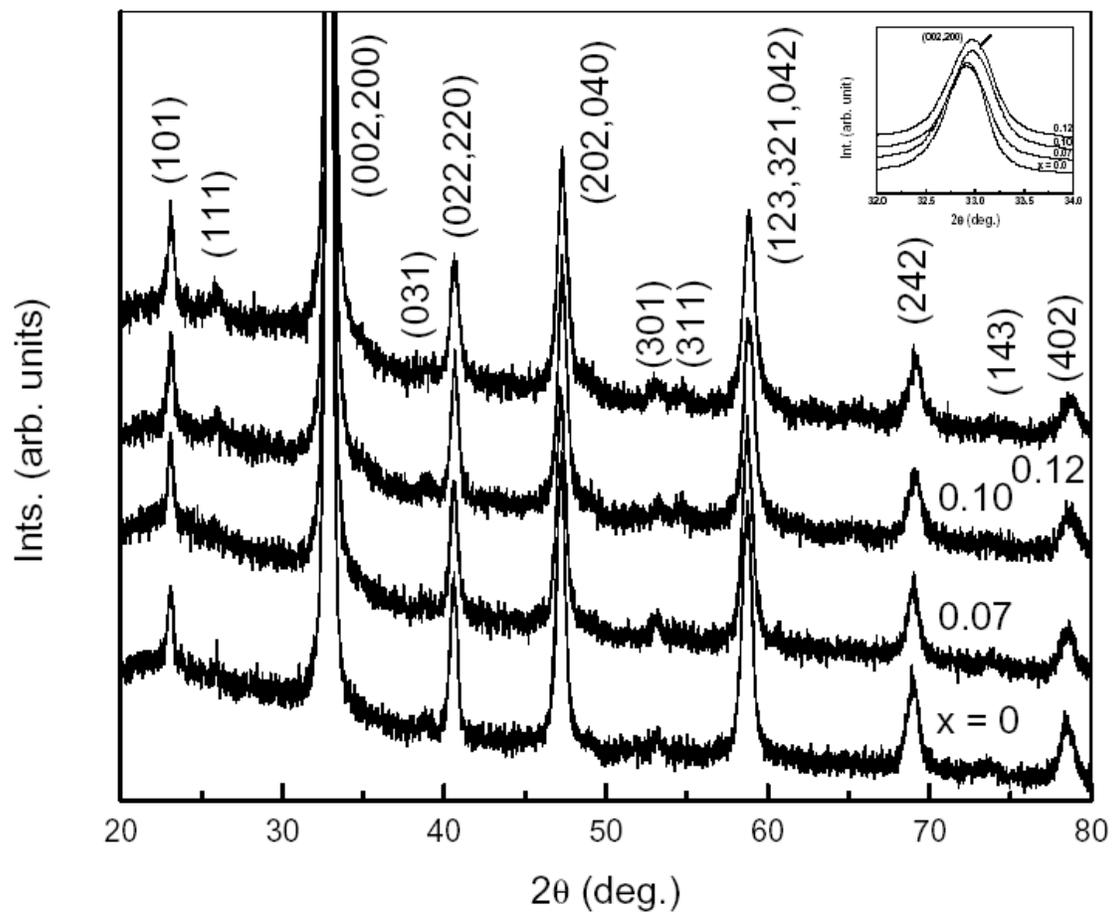

Fig. 1

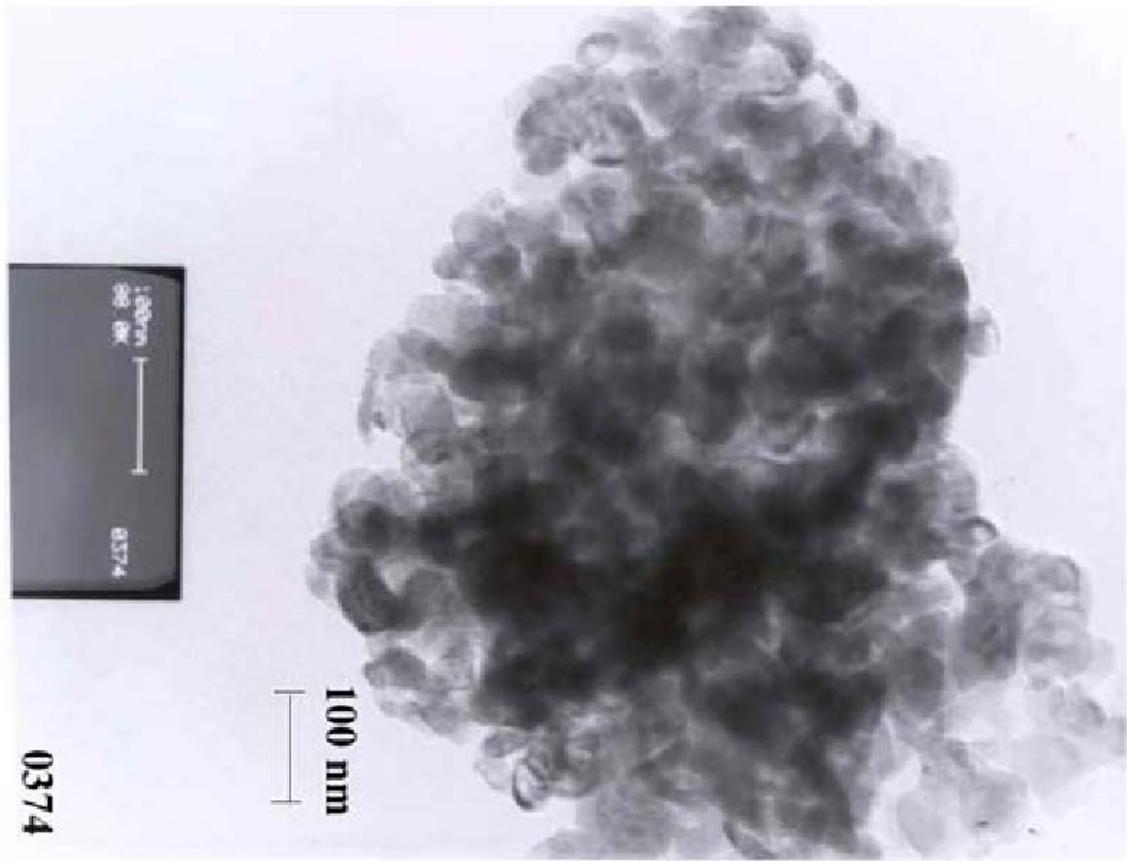

Fig. 2a

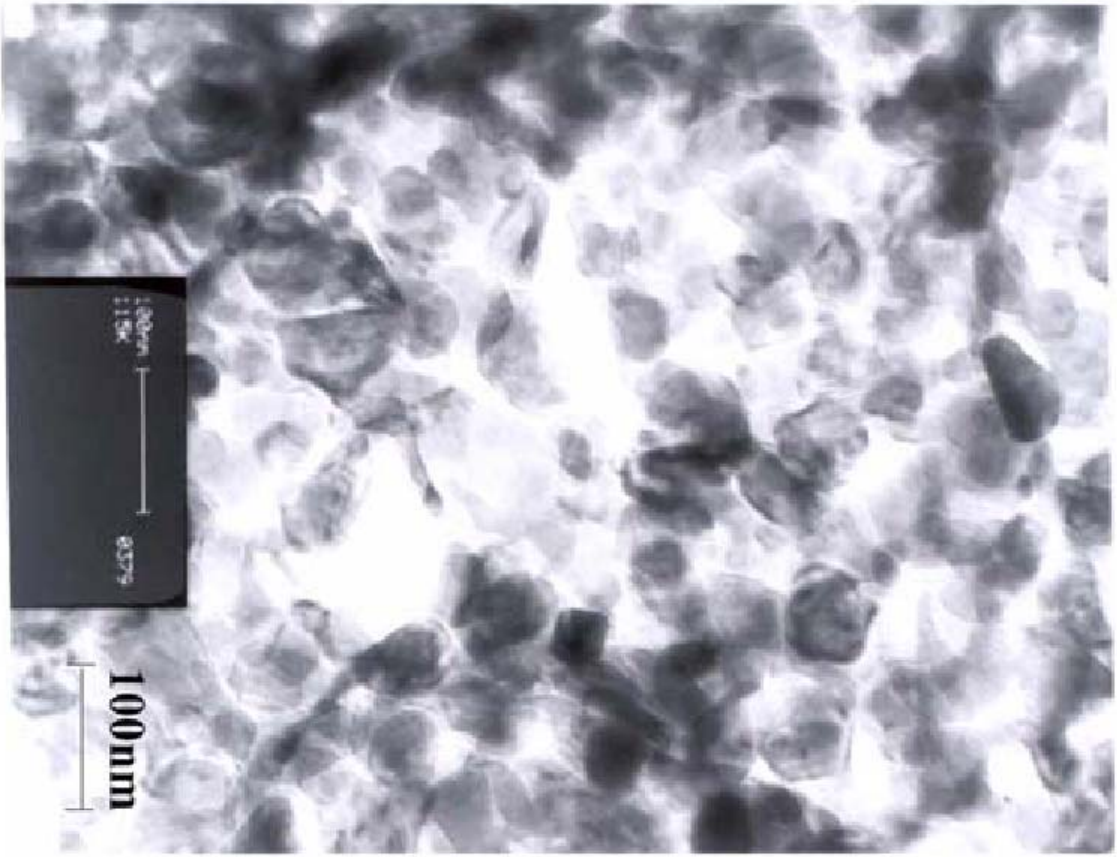

Fig. 2b

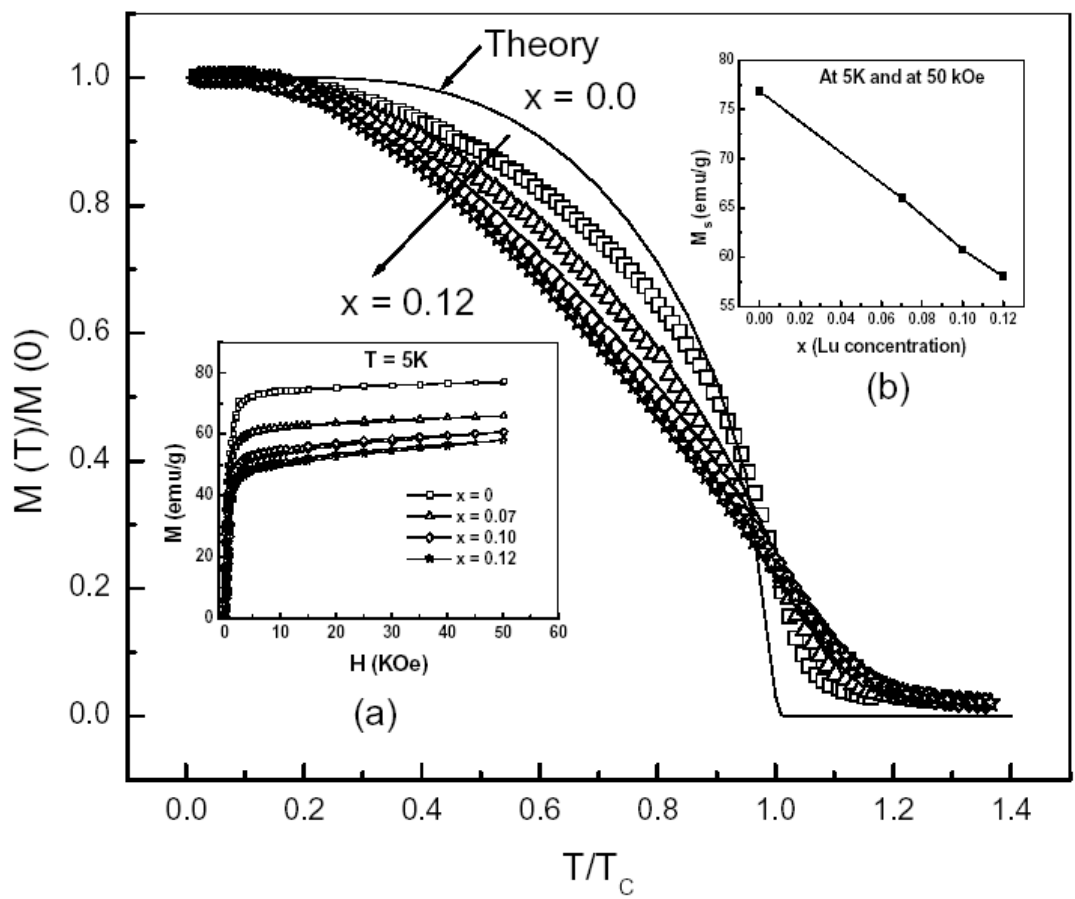

Fig. 3

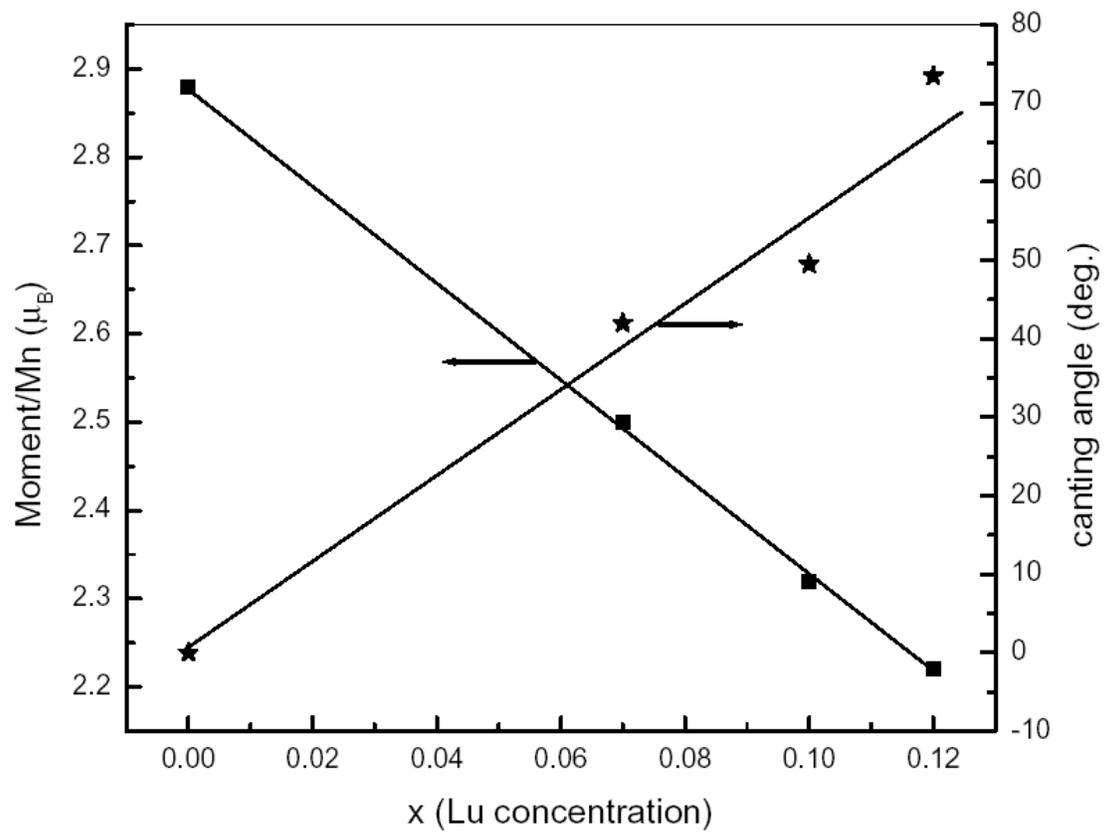

Fig. 4

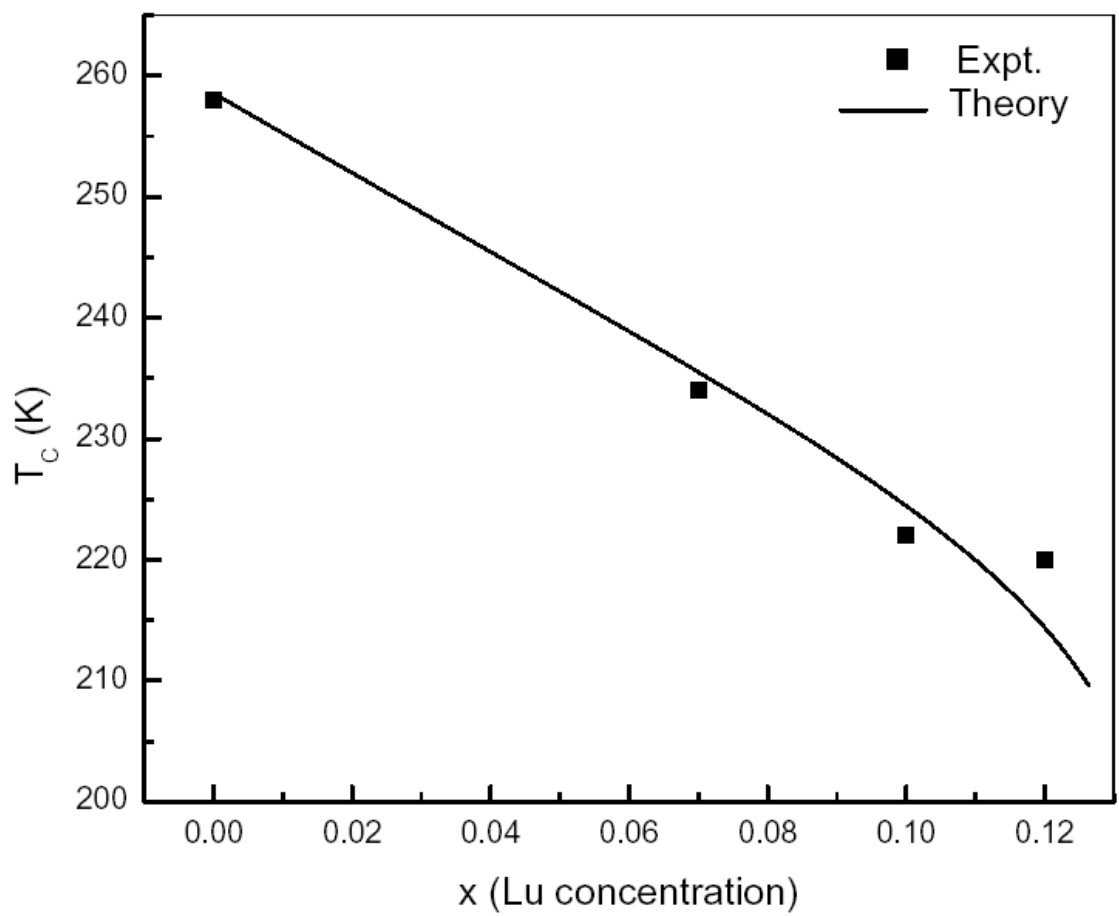

Fig. 5

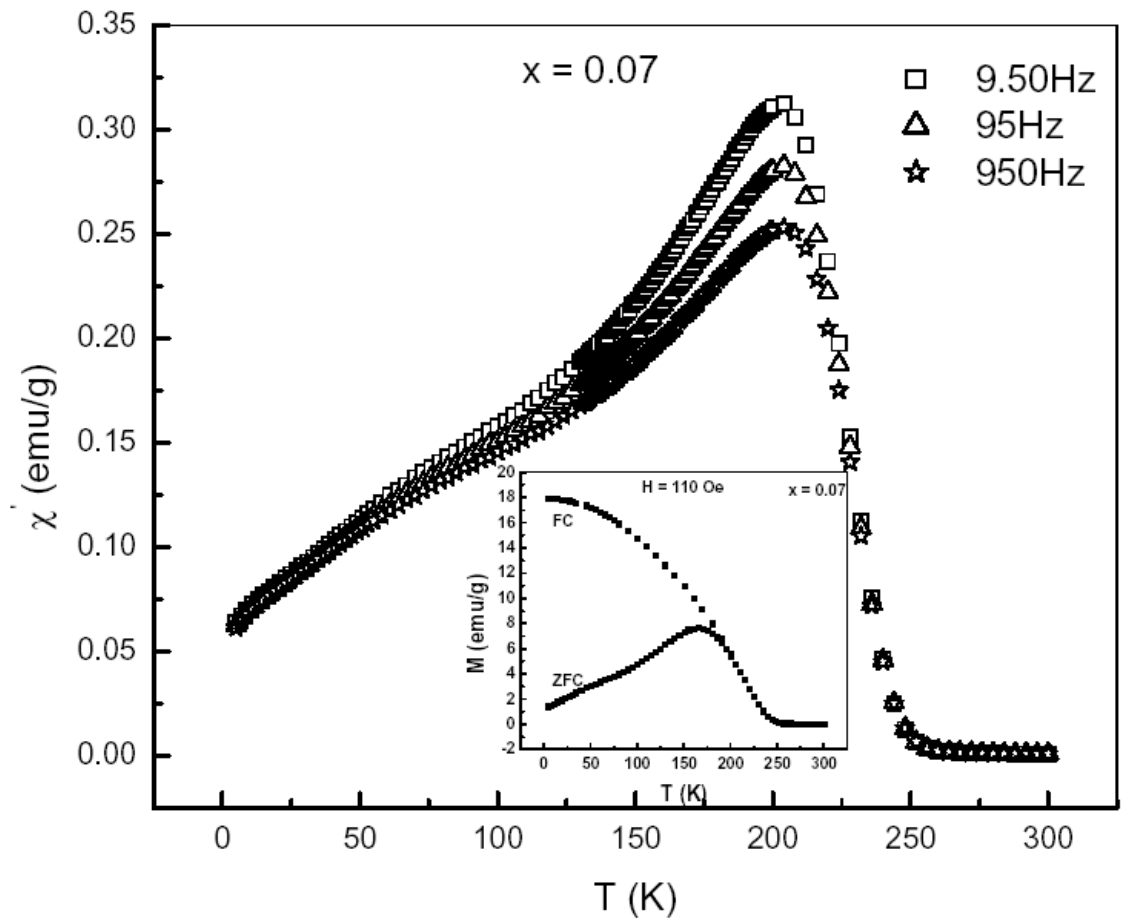

Fig. 6a

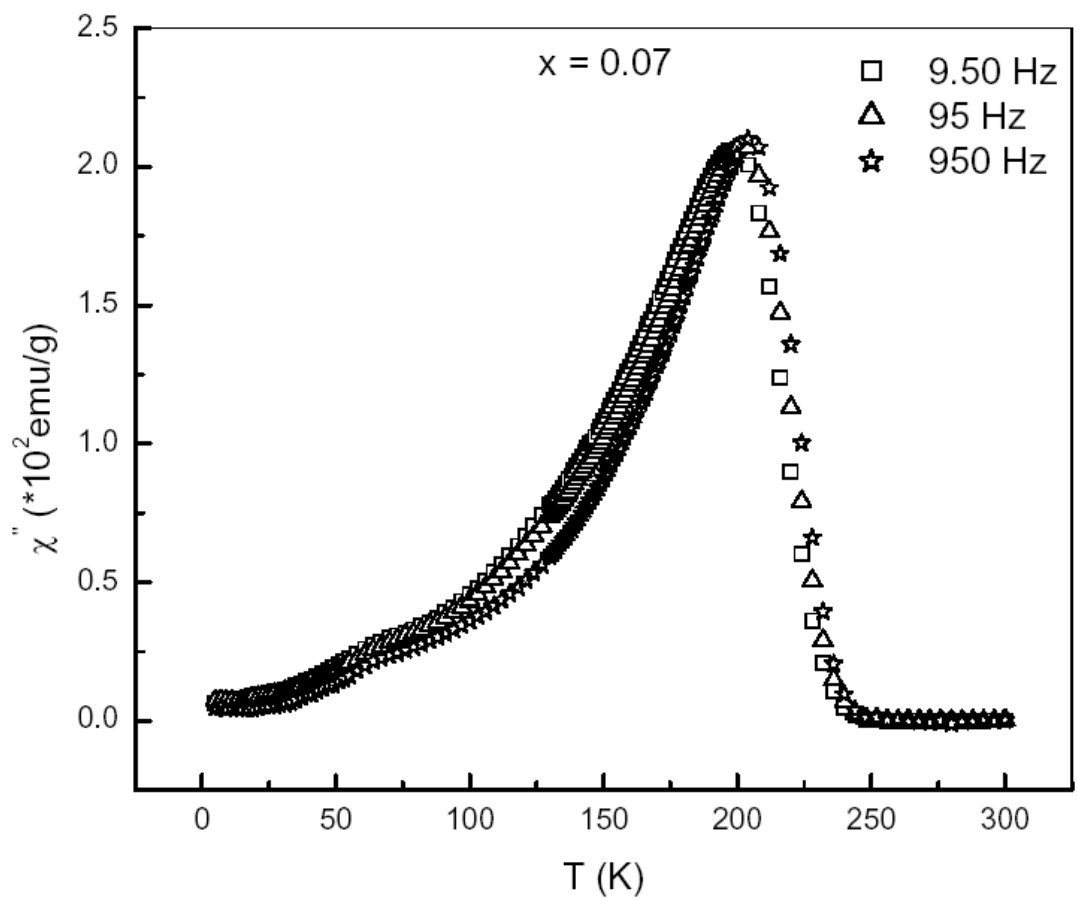

Fig. 6b

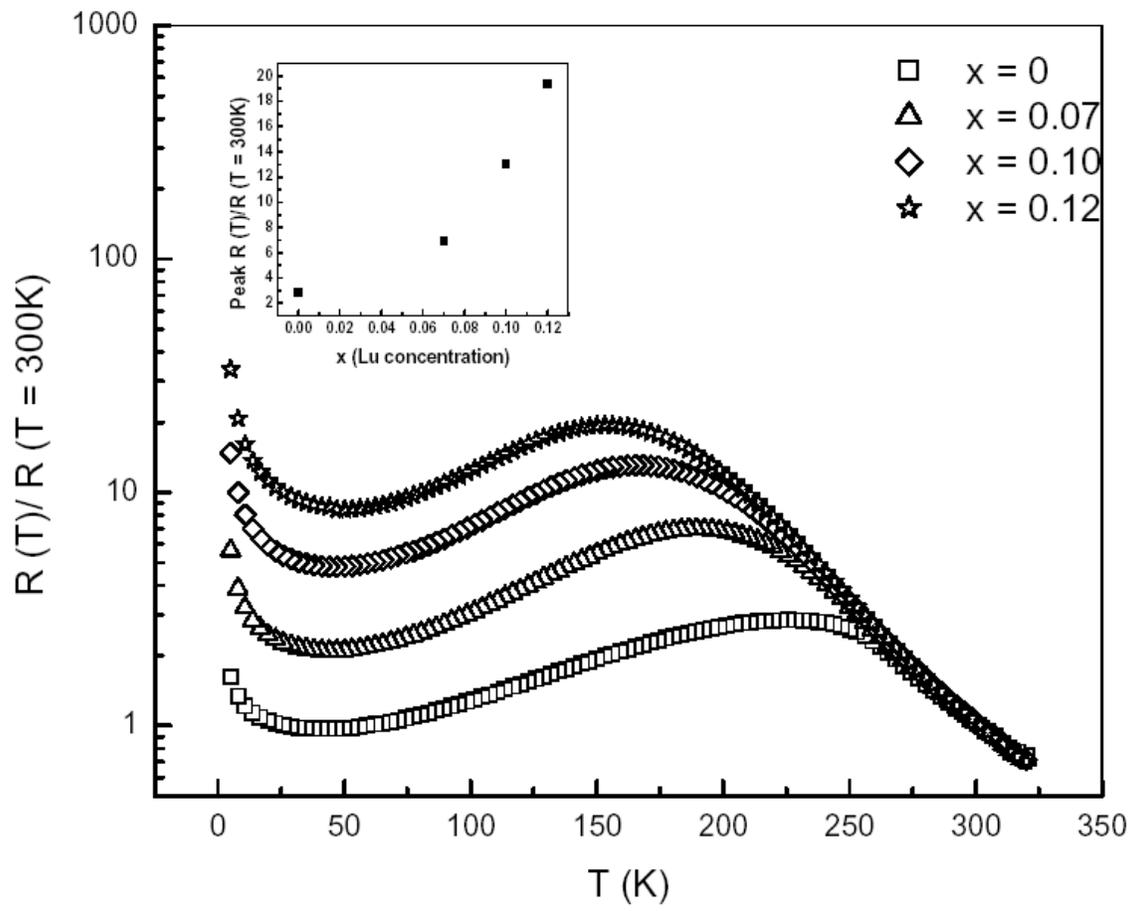

Fig. 7

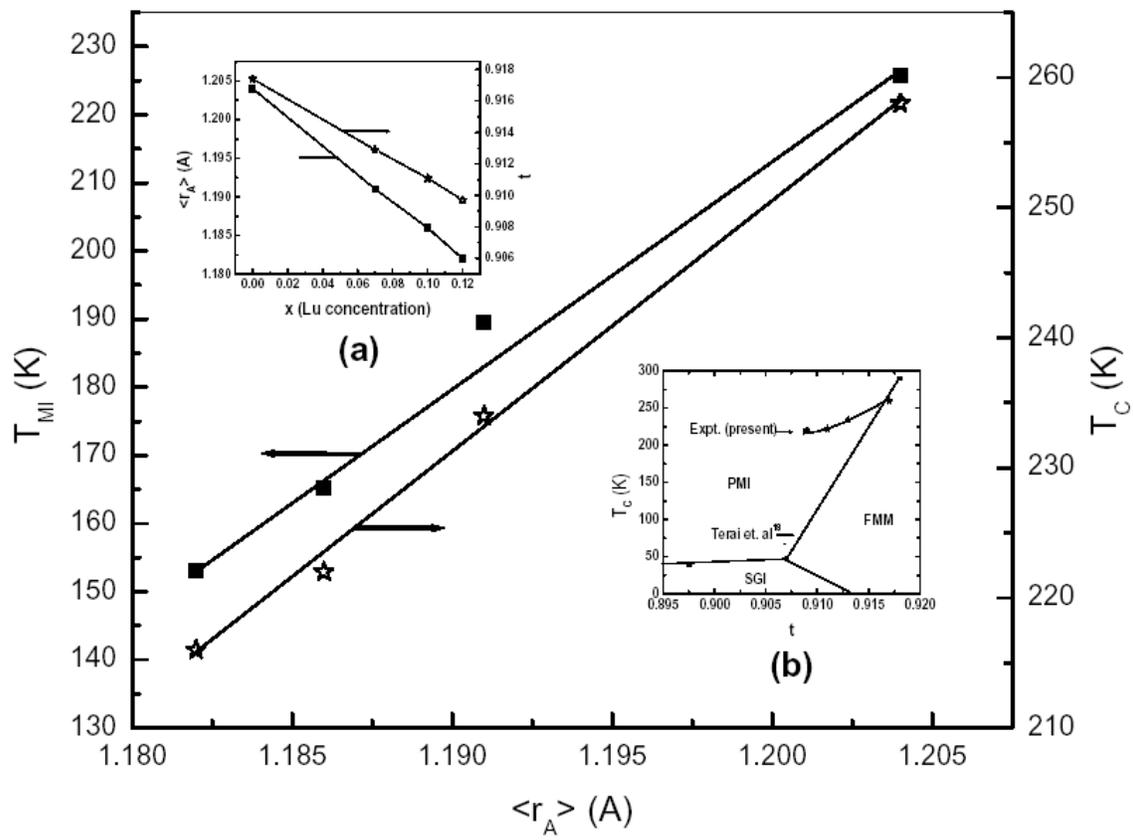

Fig. 8

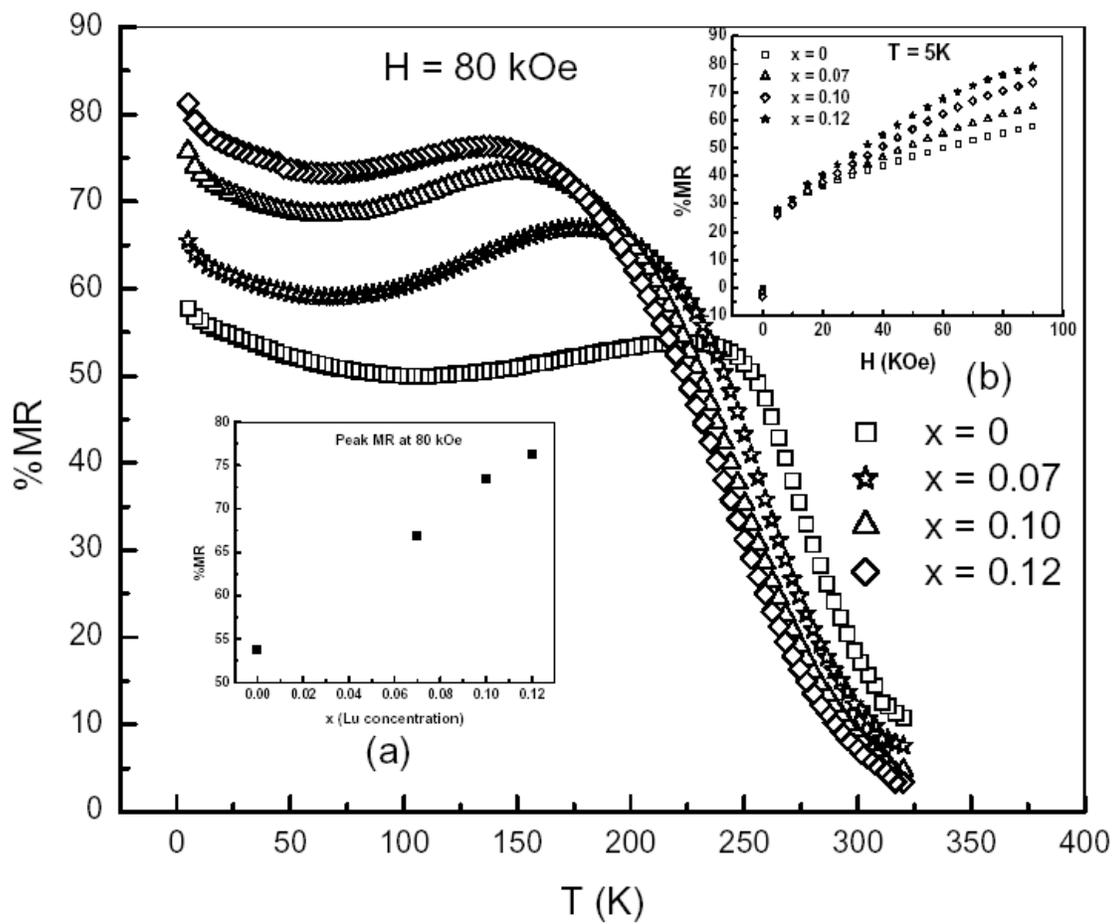

Fig. 9

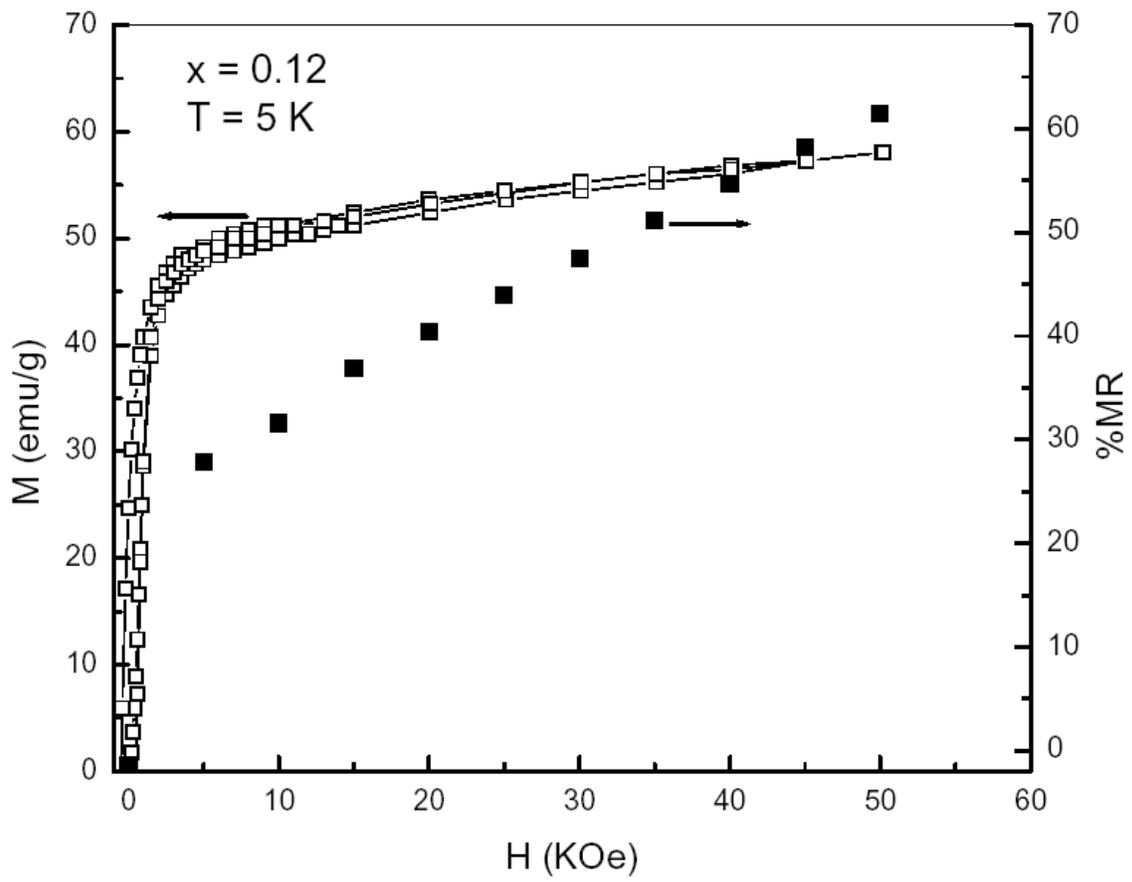

Fig. 10